\documentclass[a4paper,fleqn,usenatbib]{mnras}
\usepackage[T1]{fontenc}
\usepackage{ae,aecompl}
\usepackage{graphicx}
\usepackage{amssymb}
\usepackage{amsmath}
\usepackage{mathptmx}
\usepackage{epstopdf}
\usepackage{booktabs}
\usepackage{float}
\usepackage{subfig}
\usepackage{epsfig,color,ulem}
\usepackage{tabularx}

\definecolor{green}{rgb}{0.0, 0.5, 0.0}

\def\gsim{ \lower .75ex \hbox{$\sim$} \llap{\raise .27ex \hbox{$>$}} }
\def\lsim{ \lower .75ex\hbox{$\sim$} \llap{\raise .27ex \hbox{$<$}} }

\title[]{
Implications of the radio  and X-ray emission that followed GW170817}

\author[ Nakar \& Piran]{
	Ehud Nakar$^{1}$\thanks{udini@wise.tau.ac.il},
	Tsvi Piran$^{2}$,
	\\
	$^{1}${The Raymond and Beverly Sackler School of Physics and
		Astronomy, Tel Aviv University, Tel Aviv 69978, Israel}\\
	$^{2}${Racah Institute of Physics, The Hebrew University of
		Jerusalem, Jerusalem 91904, Israel}\\
}
\pubyear{2018}
\begin{document}
	\label{firstpage}
	\pagerange{\pageref{firstpage}--\pageref{lastpage}}
	\maketitle	
\begin{abstract}
The radio and X-rays  that followed  GW170817 are unlike any afterglow observed before, showing a gradual rise over $\sim 100$ days.They resemble the radio flare predicted long ago to follow binary neutron star mergers. This emission arises from the interaction of the merger outflow  with the external medium \citep{nakar2011}. Here we study the constraints that the observations pose on this outflow. Considering a blast wave moving  with a Lorentz factor $\Gamma$, we show that an off-axis observer, namely an observer at $\theta_{obs}> 1/\Gamma$, sees a light curve rising faster than $F_\nu \propto t^3$. Therefore, the observed rise, $F_\nu \propto t^{0.78}$,  implies that  at all times we have seen an on-axis emission. Namely,  the emitting matter was within  $\theta_{obs}<1/\Gamma$ at the time of observations (even if it was off-axis beforehand).   The observations tightly constrain the blast wave Lorentz factor:  $\Gamma \sim (1.5-7) (t/10 ~{\rm day})^{-0.21}$.  The isotropic equivalent energy in the observed region is  $E_{iso} \sim 10^{50}{\rm~erg}~ (t/150 ~{\rm day})^{1.3}$. The  energy increase can arise from a slower material moving behind the blast wave or from a matter moving at larger angles that has slowed down.  While such a structure can have different origins, the only physically motivated one proposed so far, is the interaction of a relativistic jet with the ejecta and the resulting cocoon. The jet could have been choked or it could have emerged successfully from the ejecta.  In the latter case it has produced  a short GRB pointing elsewhere (this successful jet-cocoon system is sometimes called a ``structured jet''). Although  circumstantial evidence disfavors a successful jet, the fate of the jet (choked or successful) cannot be decisively determined  from current observations. Unfortunately,  the light curve alone may not be sufficient to resolve this question, since both  chocked and  successful jets can lead to a gradual rise to a peak, followed by a decay. Therefore, the recent turnover of the light curve does not necessitate the existence of a successful jet.
\end{abstract}
\section{Introduction}

The gravitational wave radiation from the binary neutron star merger GW170817 \cite{abbott2017GW}, were followed by prompt  $\gamma$-rays\citep{Goldstein17,savchenko2017}, a  UV/optical/IR macronova/kilonova \citep{coulter2Ruan2018017,arcavi2017,soares-santos2017,diaz2017,mccully2017,buckley2018,utsumi2017,covino2017,shappee2017,Evans17,cowperthwaite2017,smartt2017,Kasliwal17,Pian17,tanvir2017,drout2017,kilpatrick2017,nicholl2017,shappee2017} and a multi-wavelengths afterglow emission, ranging from radio via optical to X-rays \citep{Troja17,Haggard2017,Margutti17,Hallinan17,Ruan2018,Alexander2017,pooley2017,mooley2018,DAvanzo18,margutti2018,lyman2018,Resmi2018,dobie2018}. { At first glance the prompt $\gamma$-rays seem to resemble a short gamma-ray burst (sGRBs) \citep{Goldstein17}. However,  a close inspection reveals that they} were different from all other sGRBs observed before \cite[e.g.,][]{Kasliwal17}. The emission that followed was also different than any GRB afterglow seen before. All GRB afterglows have been detected shortly after, or even during, the burst. While some afterglows have shown initial rapid rise, all afterglows  peak at a very early time and show later  a general decline (with some short lived plateau and flaring episodes observed mostly in long bursts).  

Even though the burst was extremely nearby, X-rays were not detected during the first nine days \citep{Evans17,Troja17} and  radio for the first sixteen  \citep{Hallinan17}. Once detected the light curve increased gradually roughly  as $t^{0.78}$ over a decade in time and  its spectrum had a 
single power-law ranging from radio to X-rays \citep{mooley2018,Ruan2018,margutti2018,lyman2018}. 
The most plausible explanation for the observed emission is synchrotron radiation from electrons accelerated at the forward shock driven by the merger outflow into the circum-merger medium.  

Here we explore, using general arguments that follow from afterglow theory \citep{sari98,sari99}, the implications of the observed emission. The radio to X-ray spectrum indicates that all the observed bands are above the self absorption and synchrotron frequencies and below the cooling frequency. We show that in this spectral regime:
\begin{itemize} 
\item[(i)] An off-axis emission, defined as emission observed from a relativistic blast wave moving with a Lorentz factor  $\Gamma$ and observed from an angle $\theta_{obs}$ satisfying: $\theta_{obs} > 1/\Gamma$ (see \S \ref{sec:term}), cannot rise slower than $t^{3}$. 
\item[(ii)] At a given time the flux from an on-axis blast wave, for which $\theta_{obs} < 1/\Gamma$ (see \S \ref{sec:term}),  depends very  strongly  on $\Gamma$, roughly as $F_\nu \propto \Gamma^{10}$. 
{We  find $\Gamma$  of the blast wave at a given time as a function of the other parameters of the system and show that it is}   constrained strongly by the observations. 
\item[(iii)] The gradual rise implies that energy is injected continuously into the blast wave. { A flattening implies a decrease at the rate of energy injection and a fast decay is expected once energy injection stops.} We quantify that and constrain the energy in the blast wave as a function of time. \end{itemize} 

These findings have several important implications on the source: 
\begin{itemize} 
\item First, the observed gradual rise in the afterglow  implies that the emitting material is moving towards us (i.e. we are within its $1/\Gamma$ cone) at least since the first X-ray detection at  day 9.  
\item Second, the energy injection  implies, in turn, that the emitting material must have  either  a radial or angular structure, or both. 
As the observed signal arises from a non-trivial convolution of the radial and angular structure of the outflow it is clear that there is a family of solutions that can fit the data. However, given the strong dependence of the signal on the Lorentz factor, one can constrain it strongly and show that $\Gamma \sim 1.5-7$, ruling out extreme relativistic or Newtonian blast waves. 
\end{itemize} 

The lack of a signal in the first few days shows that this emission did not arise from an ultra-relativistic outflow pointing towards us and it also didn't originate from an off-axis ultra-relativistic outflow pointing elsewhere at the time of observations. Instead the outflow must be mildly relativistic and on-axis at the time of observations, and most likely it was ejected at mildly relativistic velocities. This is the so called ``radio flare'' predicted long ago \citep{nakar2011}  to follow binary neutron star mergers, and elaborated on by \cite{piran2013,hotokezaka2015} and others.  These authors focused on  emission from the Newtonian ejecta moving at 0.1-0.3c that would have peaked after months to years,  but stressed that a strong early emission from a mildly relativistic component, which seems to be the emission observed here, is plausible as well.  

It seems that there is some unclarity concerning both the origin and the nature of the structure that leads to energy injection. This unclarity is linked to yet another important question, whether a jet has emerged in this case from the ejecta or not. That is whether the event was accompanied by a regular sGRB pointing elsewhere.  
Several papers have fitted  the afterglow observations to specific numerical models. 
In some of those papers the structure arose from the interaction of the jet with the merger ejecta \citep{Gottlieb17b,lazzati2017c,nakar2018b,margutti2018}. 
While different names have been given to this solution, in all those cases it is simply the cocoon that arises as the jet propagates \citep{ramirez-ruiz2002,Nakar2017,lazzati2017a} and the jet itself in case that it  breaks out, even though at times it is called a ``structured jet''.
In other papers \citep{lyman2018,DAvanzo18,Troja2018},
the structure is parameterized by a simple functional form (e.g., a Gaussian or a power-law) without any specific physical motivation apart for the fact that a jet is expected to have soem structure induced by the launching process and the interaction with the ejecta. Each of the methods (calculating the outcome of the interaction in specific setups and using an ad-hoc parametrization) has its own advantages and disadvantages, but they do not represent alternative models. We clarify this point and in particular we show that the so called ``cocoon afterglow'' arises independently of the question whether the jet has emerged from the ejecta or not.

The structure of the paper is as follows. We begin in \S \ref{sec:term} with definitions of the terminology that clarifies the framework for the discussion. We  turn to discuss afterglow properties from  on-axis and off-axis emission  in \S \ref{sec:after}.  We explore how one can infer  from the observed light curve the conditions within the emitting region putting limits on the parameters at hand.
We also show that  an off axis afterglow must rise at least as fast as $t^3$. We discuss the implications of these results to the afterglow of GW170817 in \S \ref{sec:implications} showing that we have not observed an off-axis emission and that the outflow must have structure.
We continue in \S \ref{sec:origin} with a discussion of the origin of the structure and its properties in various scenarios. In \S \ref{sec:jetNojet} we consider the question whether a jet has emerged and produced a sGRB pointing away from us and discuss what can be learned from the future evolution of the light curve in \S \ref{sec:future}. We summarize our findings in \S \ref{sec:conc}.

\section{Terminology}
\label{sec:term}

We begin by defining several key critical concepts.  We consider an emitting region that moves at a relativistic velocity with a Lorentz factor $\Gamma$.  In this case we distinguish between two types of emission. 

\noindent \underline{{\it On-axis emission}} - emission from  material (relativistic or newtonian)  where the angle between its velocity and the line-of-sight, $\theta_{obs}$, satisfies:  $ \theta_{obs} \lesssim 1/\Gamma$. Note that the emitting region does not have to move directly towards the observer, since the observed signal is insensitive to the exact angle between the emitter velocity and the line-of-sight as long as it is $ \lesssim 1/\Gamma$. Such emission from relativistic material is enhanced due to Lorentz boost.

\noindent \underline{{\it Off-axis emission}} - emission from relativistic material with a Lorentz factor $\Gamma$ where the angle between the emitter velocity and the line-of-sight satisfies $\theta_{obs} >1/\Gamma$. The observed emission is significantly fainter than that observed by an on-axis observer of the same emission, except in the case of low $\Gamma$. 

Since the emission at any given time is largely independent of the history of the emitting region the distinction between the two categories is {\it time dependent}. Namely, a source that is "off-axis" relative to a specific observer at one time becomes "on-axis" at a later time once it decelerates and expands sideways. Clearly, since the blast wave does not accelerate, the opposite cannot happen. A source that it initially "on-axis" remains "on-axis" at all times. 

We turn now to definitions that concern the  model of the emitting region. 

\noindent \underline{{Structured relativistic jet}} - a generic name for any outflow with an angular, and possibly  a radial, structure where along the symmetry axis there is a relativistic jet. This, by itself, is not a physical model until the specific structure (i.e., energy and velocity distributions as a function of the angle) is defined. 
One specific physically motivated model for a structured jet is a successful jet with a cocoon (see below). 

\noindent \underline{{Cocoon}} - The outflow that is formed during the propagation of a jet within a dense medium (in our case the merger ejecta). It is composed of all the ejecta material that was shocked by the blast wave driven by the jet propagation, and of all the jet material that was shocked by the reverse shock that is formed at the jet-ejecta interaction layer (known as the head of the jet). The properties of the cocoon and its formation are discussed in length in \cite{Nakar2017}.  We stress that the cocoon is an inevitable outcome of the propagation of a relativistic jet in the ejecta, and given the ejecta mass and velocity inferred from the optical emission \cite[e.g.,][]{Kasliwal17}, it must contain a considerable energy.  A cocoon is formed with either a choked jet or a successful one, as discussed below.  We also stress that the cocoon has both a radial and an angular structure which depends on whether the jet is collimated or not and on whether it is choked or not (see figures \ref{fig:choked} and \ref{fig:successful} below and \citealt{Nakar2017,Gottlieb17b,Gottlieb17a}). 

\noindent \underline{{A cocoon with a choked jet - no sGRB}} - The jet that drives the cocoon is choked before it breaks out of the ejecta. In this case the jet deposits all its energy in the cocoon and the resulting outflow is composed only of shocked jet and shocked ejecta material. Although there is no ultra relativistic jet at the core of the outflow, it does have an angular and a radial velocity structure. The outflow may include mildly relativistic material and possibly even relativistic one, depending on the ejecta initial velocity  profile and the location within the ejecta where the jet is choked. The maximal Lorentz factor, however, is not expected to be high enough to produce a short GRBs. Nevertheless, the mildly relativistic outflow may produce a different type of gamma-ray flare upon the breakout of the cocoon forward shock from the ejecta, which may very well be the gamma-ray signal observed in GW170817 \citep{Kasliwal17,Gottlieb17b,nakar2018b}.

\noindent \underline{{A cocoon with a successful jet - sGRB pointing along the jet axis}} - The jet that drives the cocoon does break out. After breakout  there is no more ejecta material at the head of the jet and it can propagate at ultra relativistic velocity to infinity (or until it interacts with the ISM).  An observer that is within the opening angle of the jet will see a regular sGRB. The cocoon breaks out along with the jet. { Also in this case the shock breakout of the cocoon produces a low-luminosity wide angle gamma-ray flare \citep{Gottlieb17b,bromberg2018,Pozanenko2018,nakar2018b}.}  After breakout the jet and the cocoon decouple in the sense that there is no more significant energy deposition into the cocoon and all (or at least most) of the luminosity that is launched into the jet at its base remains in the form an ultra-relativistic jet. In this scenario the resulting outflow is an ultra-relativistic jet at the core which is surrounded by a mildly relativistic and Newtonian cocoon material with an angular and a radial structure (see e.g., \citealt{Gottlieb17a}). A cocoon with a successful jet is a structured jet. In fact it is the only physically motivated structured jet model that was suggested in the context of the afterglow of GW170817. In the broader context of GRB afterglows, as far as we know, it is the only structured jet model that is based on a physical study of the interaction between the jet and the ejecta, which must take place in every GRB.   

\section{Inferring the outflow properties from the afterglow emission}
\label{sec:after}	
In all the models we consider here the radio and X-ray emission is generated by optically thin synchrotron  from an ISM material that was heated via the forward shock driven by the merger outflow (ejecta, cocoon and possibly jet). This origin for the emission is supported by the observed spectrum \citep[both in the individual radio and X-ray bands and in the radio to X-ray flux ratio;][]{mooley2018,margutti2018,DAvanzo18}. Given that we observe a single power-law from radio to X-ray with $F_\nu \propto \nu^{-0.6}$, we conclude that the whole observed spectrum is within a single segment of the synchrotron spectrum:  below the cooling frequency, $\nu_c$, and above the typical synchrotron frequency, $\nu_m$, and the self absorption frequency $\nu_a$. 
\subsection{On-axis emission}
We consider a quasi-spherical on-axis outflow, namely the outflow is spherical over an angle of $1/\Gamma$ with respect to the line-of-sight, where $\Gamma$ is the Lorentz factor of the region that dominates the observed emission. {We use the standard afterglow theory \citep{sari98} to provide equations that enable to extract the blast wave energy and Lorentz factor directly from the observations. While the derivation of this useful result is simple and parts of it were explored in the context of the X-ray plateau in GRB afterglows \citep[e.g., ][]{granot2006}, as far as we know it was not presented in  this form and therefore we repeat the derivation here.}

For an on-axis observer the radius of the blast wave, $R$, at a given observer time is $R \propto \Gamma^2 t$, where $t$ is the observed time since the merger. The number of emitting electrons is $N_e \propto R^3 \propto \Gamma^6 t^3$. The flux at $\nu_m$ satisfies $F_{\nu,m} \propto \Gamma B N_e  \propto \Gamma^8 t^3$ and $\nu_m \propto B\Gamma^3 \propto\Gamma^4$, where $B \propto \Gamma$ is the magnetic field in the emitting region. The flux in the spectral range $\nu_m,\nu_a<\nu<\nu_c$ satisfies $F_\nu \propto F_{\nu,m} \nu_m^{\frac{p-1}{2}} \propto \Gamma^{6+2p}$, where $p$ is the power-law index of the electron distribution. This extremely strong dependence of the observed flux on $\Gamma$ implies that observations of an on-axis emission tightly constrain the blast wave Lorentz factor at any time. 

When the entire set of parameters are considered we obtain
\begin{eqnarray}
	F_{\nu,on-axis} \approx 20\mu Jy~ \left( \frac{\Gamma}{4}\right)^{6+2p} \left( \frac{n}{10^{-4}{\rm~cm^{-3}}}\right)^\frac{p+5}{4}
	\\ \nonumber
	\epsilon_{e,-1}^{p-1} \epsilon_{B,-2}^\frac{p+1}{4}
	\left(\frac{t}{10d}\right)^3 \left( \frac{\nu_{obs}}{3GHz}\right)^{-\frac{p-1}{2}}  \left(\frac{d}{40Mpc}\right)^{-2} \ , 
\label{eq:Fnu} \end{eqnarray}
where $\epsilon_{e,B}$ are the standard equipartition parameters of the electrons and the magnetic field, $n$ is the external density (assumed implicitly to be uniform) and $d$ is the source distance.  
Inverting this equation we can determine the 
blast wave  Lorentz factor, $\Gamma$,  at time $t$:\begin{eqnarray}
	\Gamma (t) \approx 4 ~ \left(\frac{F_\nu}{20\mu {\rm Jy}} \right)^{\frac{1}{6+2p}} 
	\left( \frac{\nu_{obs}}{3GHz}\right)^{\frac{p-1}{12+4p}} 
	\left(\frac{t}{10d}\right)^{-\frac{3}{6+2p}} \nonumber \\
	\epsilon_{e,-1}^{-\frac{p-1}{6+2p}} \epsilon_{B,-2}^{-\frac{p+1}{24+8p}}
	\left( \frac{n}{10^{-4}{\rm~cm^{-3}}}\right)^\frac{-(p+5)}{24+8p} \left(\frac{d}{40Mpc}\right)^\frac{1}{3+p} \ .
	\label{eq:gammaon}
\end{eqnarray}

From the evolution of the flux with time,  we can infer the evolution of $\Gamma$ with time and from this the isotropic equivalent blast wave energy within a region of size $1/\Gamma$, (or a solid angle $\sim \pi/\Gamma^2$), around the line of sight:  
\begin{eqnarray}
E_{iso}(t) 
 \approx   5 \times 10^{48}~{\rm erg}~ \left( \frac{F_\nu}{20\mu Jy}\right)^\frac{4}{3+p} 
 \left( \frac{\nu_{obs}}{3GHz}\right)^{\frac{2(p-1)}{3+p}}
\left(\frac{t}{10d}\right)^{\frac{3(p-1)}{3+p}} \nonumber  \\
 \left( \frac{n}{10^{-4}{\rm~cm^{-3}}}\right)^{-\frac{2}{3+p}} %
	\epsilon_{e,-1}^{-\frac{4(p-1)}{3+p}} \epsilon_{B,-2}^{-\frac{p+1}{3+p}} \left(\frac{d}{40Mpc}\right)^\frac{8}{3+p}\ .
\label{eq:Eoon}
\end{eqnarray}   
For a constant $E_{iso}$  we recover the well known regular afterglow  light curves in this spectral regime \citep{sari98}. The total energy in the observed region, $E_{obs}$, is $\sim E_{iso}/2\Gamma^2$, namely:
\begin{eqnarray}
E_{obs}(t) 
 \approx   2 \times 10^{47}~{\rm erg}~ \left( \frac{F_\nu}{20\mu Jy}\right)^\frac{3}{3+p} 
 \left( \frac{\nu_{obs}}{3GHz}\right)^{\frac{3(p-1)}{2(3+p)}}
\left(\frac{t}{10d}\right)^{\frac{3p}{3+p}} \nonumber  \\
 \left( \frac{n}{10^{-4}{\rm~cm^{-3}}}\right)^{-\frac{3-p}{4(3+p)}} %
	\epsilon_{e,-1}^{-\frac{3(p-1)}{3+p}} \epsilon_{B,-2}^{-\frac{3(p+1)}{4(3+p)}} \left(\frac{d}{40Mpc}\right)^\frac{6}{3+p}\ .
\label{eq:Eoonobs}
\end{eqnarray}

For a power-law temporal evolution of the observed flux, $F_\nu =  F_{\nu_,0}  (t/t_0)^{\alpha}$,  Eqs. \ref{eq:gammaon} and \ref{eq:Eoon} dictates: $\Gamma \propto t^{-\frac{3-\alpha}{6+2p}}$ and $E_{iso} \propto t^{\frac{4\alpha + 3(p-1) }{3+p}}$.  Combining the last conditions  we can obtain the isotropic equivalent energy within the observed region (of angular size $o(\Gamma^{-2}$)) when the blast waves moves at a Lorentz factor $\Gamma$: 
\begin{eqnarray}
E_{iso}(>\Gamma) 
 \approx   5 \times 10^{48}~{\rm erg}~ 
 \left( \frac{\Gamma}{4}\right)^{-\frac{8\alpha+6(p-1)}{3-\alpha}}
 \left( \frac{F_{\nu,10d}}{20\mu Jy}\right)^\frac{3}{3-\alpha} 
 \left( \frac{\nu_{obs}}{3GHz}\right)^{\frac{3(p-1)}{6-2\alpha}}
 \nonumber  \\ 
 \left( \frac{n}{10^{-4}{\rm~cm^{-3}}}\right)^{-\frac{3(p+1)+4 \alpha}{4(3-\alpha)}} 
	\epsilon_{e,-1}^{-\frac{3(p-1)}{3-\alpha}} \epsilon_{B,-2}^{-\frac{3(p+1)}{4(3-\alpha)}} \left( \frac{D}{40Mpc}\right)^\frac{6}{3-\alpha} \ .
\label{eq:EGamma}
\end{eqnarray} 
The total energy in the observed region as a function of $\Gamma$ is simply $E_{obs}(>\Gamma)\approx E_{iso}(>\Gamma)/2\Gamma^2$. 
These equations hold as long as the blast wave is relativistic. In the Newtonian regime, \cite{piran2013} find
$E(>\beta) \propto (\beta)^{-\frac{15p+10\alpha-21}{6-2\alpha}}$.

\subsection{Off-axis emission}
The calculations of the off-axis case are  more complicated as they depend on additional parameters:
the viewing angle and the opening angle of the jet as well as on the details of the sideways expansion. However, here our goal is more modest. We obtain a lower limit on how fast the  light curve of an off-axis emission should rise.
Consider an element of the blast wave that is moving at an angle larger than $1/\Gamma$ with respect to the observer.  Assume 
for simplicity that it is not expanding sideways, so that it occupies a constant solid angle and moves at a constant angle to the observer. We 
have to modify Eq. \ref{eq:Fnu} to account for the different boosts and the different relation between the observer 
time and the blast wave time. Now $N_e \propto R^3 \propto t^3$ and the Doppler boost is $ \Lambda \propto 1/\Gamma$ while 
the magnetic field $B \propto \Gamma$, implying $F_{\nu,m} \propto N_e B \Lambda^{3}\propto \Gamma^{-2} t^3$ and $\nu_m 
\propto \Gamma^2 B \Lambda \propto \Gamma^2$. Thus, the flux density at a given frequency evolves as  $F_{\nu,off-axis} \propto 
F_{\nu,m} \nu_m^{\frac{p-1}{2}} \propto \Gamma^{-(3-p)} t^3$. 

If there is no lateral spreading $\Gamma \propto t^{-3/2}$ in the deceleration phase and $\Gamma \propto {\rm const.}$  in the coasting phase.   In reality we expect the jet to spread sideways, and the flux to rise even faster. Therefore, the observed flux from an off-axis emitter must rise faster than $t^3$.
This conclusion is independent of the exact structure of the relativistic material, and it is not restricted to the so called ``top hat''  jet, as long as its radiation is off-axis, namely the angle between the observer and the radiating material is much larger than $1/\Gamma$.
 In fact, the way to produce a slowly rising (or even declining) light curve from an off-axis emission is by accelerating the blast wave or making the jet narrower, both are unexpected in any realistic scenario.

\section{Implications} 
\label{sec:implications}

We turn now to explore the implications of the above results to the radio, optical and X-ray afterglow of GW 170817. 	
The radio afterglow shows a continuous rise in the flux since the first detection on day  16 and until day 115.
The rise is consistent with a single power law \citep{mooley2018}:
\begin{equation}
F_\nu \approx 13 \pm 0.4 ~\mu {\rm Jy} ~  (t/10~ {\rm days})^{0.78 \pm 0.05} ~(\nu/3 {\rm ~GHz})^{0.61 \pm 0.05} \ . 
\label{eq:obs}
\end{equation} 
{ After day 115 the light curve starts to flatten \citep{DAvanzo18,margutti2018}, reaching a peak around  day 150 \citep{dobie2018}. The radio, optical and X-rays are consistent with being on the same power-law segment, 
$\nu_a, ~ \nu_m < \nu_{r,o,x} < \nu_c$, at least until day 150 \citep{margutti2018}.}

The moderate rise in the flux, $F_{\nu} \propto t^{0.78}$,  implies that at all times the emission cannot be dominated by an off-axis emission, which must rise faster than $t^3$. 
As off-axis emission is ruled out we turn to on-axis emission to explain the observed light curve. Being on-axis we can use the quasi-spherical approximation explored earlier. 
The outflow is probably not fully quasi-spherical, and therefore this approximation provides only an order of magnitude estimate, yet the very strong dependence of $F_\nu$ on $\Gamma$ implies that the latter  can be tightly constrained. 
It  follows from the spectral index of the afterglow of GW170817 that $p \approx 2.2$ and this yields:
\begin{eqnarray}
	\Gamma(t)\approx 4~ \left( \frac{F_\nu}{20\mu Jy}\right)^{0.1} \left( \frac{\nu_{obs}}{3GHz}\right)^{0.06} \left(\frac{t}{10d}\right)^{-0.29}  \nonumber \\
	 	\epsilon_{e,-1}^{-0.12} \epsilon_{B,-2}^{-0.08} \left( \frac{n}{10^{-4}{\rm~cm^{-3}}}\right)^{-0.17} \left(\frac{d}{40Mpc}\right)^{0.19} \ .	
\end{eqnarray}
Using this equation and allowing the two most uncertain parameters to vary, $n$ in the range $0.01-10^{-5} {\rm~cm^{-3}}$ and $\epsilon_B$ in the range $0.01-10^{-4}$, we can constrain the blast wave Lorentz factor to be in the range $\Gamma \approx 1.5-7$ at the time of the first detection. The emitting region cannot be extremely relativistic and it cannot be Newtonian either.

The observed  rise implies that the Lorentz factor of the radiating material decreases with time as  $\Gamma \propto t^{-0.21}$ 
while the isotropic equivalent energy in the emitting region { between day 10 and 150 is}:
\begin{eqnarray}
E_{iso}(t) 
 \approx   10^{50}~{\rm erg}~ \left( \frac{F_{\nu,150d}}{100\mu Jy}\right)^{0.77} 
 \left( \frac{\nu_{obs}}{3GHz}\right)^{0.46}
\left(\frac{t}{150d}\right)^{1.3} \nonumber  \\
 \left( \frac{n}{10^{-4}{\rm~cm^{-3}}}\right)^{-{0.38}} %
	\epsilon_{e,-1}^{-{0.92}} \epsilon_{B,-2}^{-{0.61}} \left(\frac{d}{40Mpc}\right)^{1.53}\ ,
\label{eq:Eisop}\end{eqnarray}
and the total energy in the observed region { in the time period is}:
\begin{eqnarray}
E_{obs}(t) 
 \approx  2 \times 10^{49}~{\rm erg}~ \left( \frac{F_{\nu,150d}}{100\mu Jy}\right)^{0.58} 
 \left( \frac{\nu_{obs}}{3GHz}\right)^{0.35}
\left(\frac{t}{150d}\right)^{1.73} \nonumber  \\
 \left( \frac{n}{10^{-4}{\rm~cm^{-3}}}\right)^{-{0.03}} %
	\epsilon_{e,-1}^{-{0.69}} \epsilon_{B,-2}^{-{0.46}} \left(\frac{d}{40Mpc}\right)^{1.15}\ .
\label{eq:Etotp}\end{eqnarray}
{ where we set the canonical values to 150 days, when the flux stopped rising.} Note the very weak dependance of $E_{obs}$ on the external density. Given the energy dependence on the microphysical parameters it can vary by an order of magnitude or even more for reasonable values of  $\epsilon_B$. Clearly, $E_{obs}(t)$ is a lower limit on the total energy of the blast wave.  
{ Equation \ref{eq:Eoon} shows that although the flux stopped rising around day 150, the energy injection did not necessarily stopped at that time.}  If the flux will  decay at a rate that is similar to a regular on-axis afterglow after the jet break ($\propto t^{-p}$ in the relativistic regime and $\propto t^{-(15p-10)/21}$ in the newtonian regime; \citealt{piran2013}) it will imply that we have seen most of the outflow energy. If not, then equation \ref{eq:Eoon} can be used to estimate the energy injection rate, as long as the blast wave is mildly relativistic.  It can be easily modified to account for the blast wave energy once it becomes Newtonian.  

Finally we can estimate the dependence of the energy on the Lorentz factor { during the rising phase ($10<t<150$ day)}: 
$E(>\Gamma) \propto    {\Gamma}^{-6.2}$ 
 in the relativistic regime, while in the Newtonian regime $E(>\beta) \propto (\beta)^{-4.5}$ \citep{piran2013}. In the mildly relativistic regime \cite{mooley2018} find that $E(>\Gamma\beta) \propto (\Gamma\beta)^{-5}$ provides a good fit to the data.
This energy increase can be obtained either by a radial structure and/or by an angular structure and there is an infinite number of solutions that will give rise to the observed light curve \citep[see e.g.][]{Troja2018}.  For example, even under the assumption of a spherical symmetry, one can find a purely radial energy distribution that will fit almost any light curve, and in particular the observed one.  In reality  it is most likely that the structure is both radial and angular and both factors contribute. In a radial structure, slower moving material carries more energy than faster moving material and it follows behind it. In an angular structure the energy carried by the outflow increases towards the axis (away from the observer) and as the outflow decelerates the beam of emission from this more energetic material includes the observer (namely, it becomes an on-axis emission). Note that also in this case the emission is always dominated by on-axis emission. Namely the emission from each element in the outflow is negligible until it decelerates enough to become an on-axis emitter. 

\subsection{The origin and the properties of the outflow structure}\label{sec:origin}
The current afterglow observations  do not require a jet at all. Instead they can be produced by the fast  tail of the sub-relativistic ejecta \citep{Hallinan17,mooley2018,Hotokezaka2018}. In that case the outflow structure is mostly radial and its origin is, most likely, the dynamics of the merger itself (mostly the first impact between the two neutron stars, see \citealt{Kyutoku2014,Kiuchi2017,Bovard2017}). However, the gamma-rays require a mildly or fully relativistic source \citep{Gottlieb17a} and there seems to be no clear mechanism
in this scenario  that can produce them.

If the afterglow is driven by the power deposited by a jet, then the outflow structure is  induced by the internal structure of the jet upon launch and the outcome of the interaction with the ejecta. The effect of the initial jet structure on the final outflow structure was not investigated in detail, however the highly pressurised cocoon that form during the interaction with the ejecta is expected to filter out any low-energy wings if such are launched by the source of the jet. The structure that is induced by the interaction of the jet with the ejecta was studied numerically for some specific configurations expected following a neutron star binary merger, both before  \citep{Gottlieb17a} and after \citep{Kasliwal17,Gottlieb17b,lazzati2017b,bromberg2018,nakar2018b,margutti2018} the detection of GW170817. In all these studies the injected jet has a simple structure (either a top hat or a gaussian) but the final outflow structure is dominated by the cocoon and the jet if the latter breaks out. Below we discuss briefly some of the characteristics of this structure in various scenarios. 

\begin{figure}	
\includegraphics[width=0.5\textwidth]{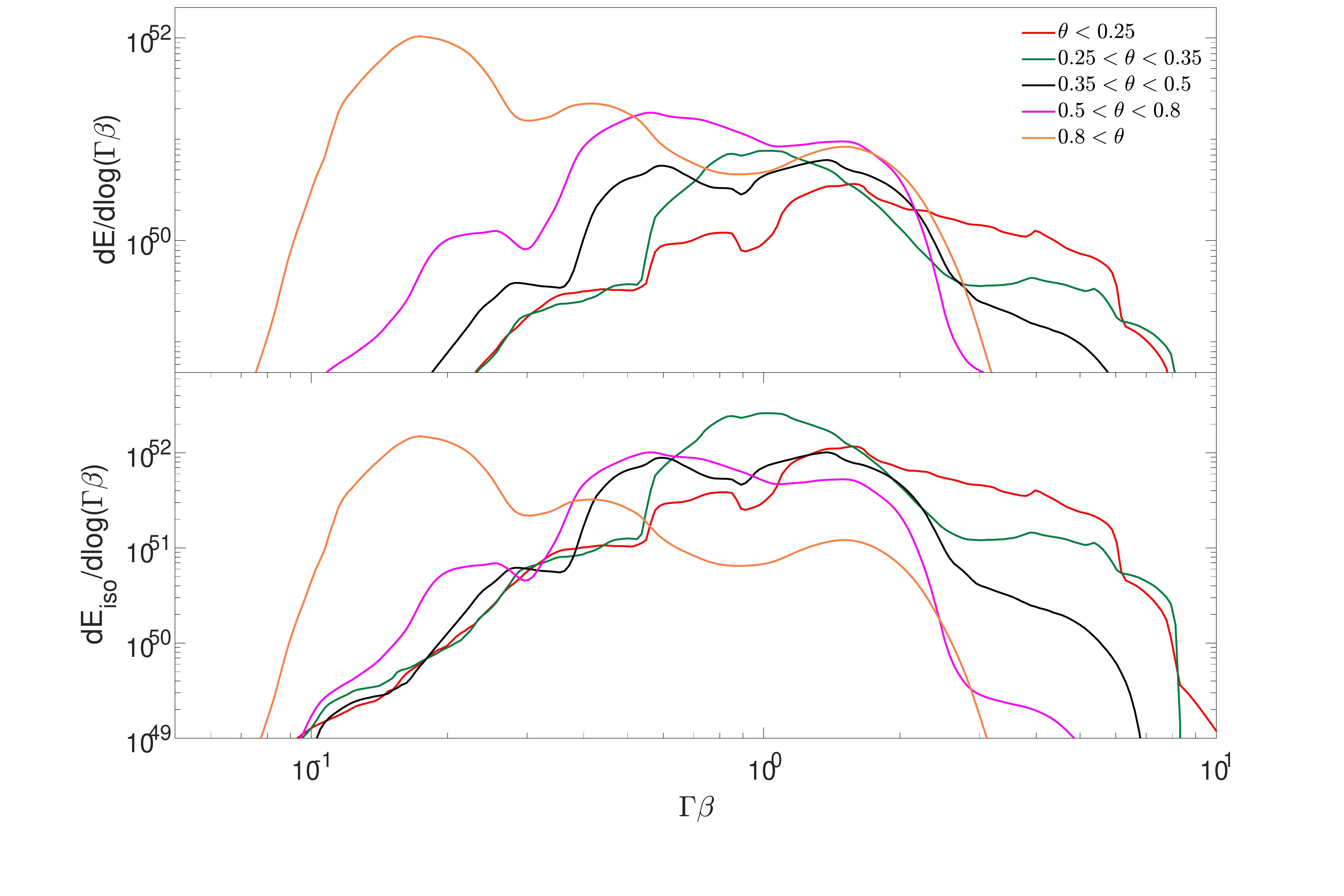}
\caption{A choked jet structure: the total energy (top panel) and isotropic equivalent energy (bottom panel) carried by the outflow per logarithmic scale of the four velocity at different angles with respect to the jet axis. The outflow is calculated by a hydrodynamic simulation of an uncollimated choked jet that was presented in  \citet{Gottlieb17b}. The figure shows that the outflow has a considerable radial and angular structure, where the material  along the axis carries much more isotropic equivalent energy at higher Lorentz factor. }
\label{fig:choked}%
\end{figure}

\begin{figure}	
\includegraphics[width=0.5\textwidth]{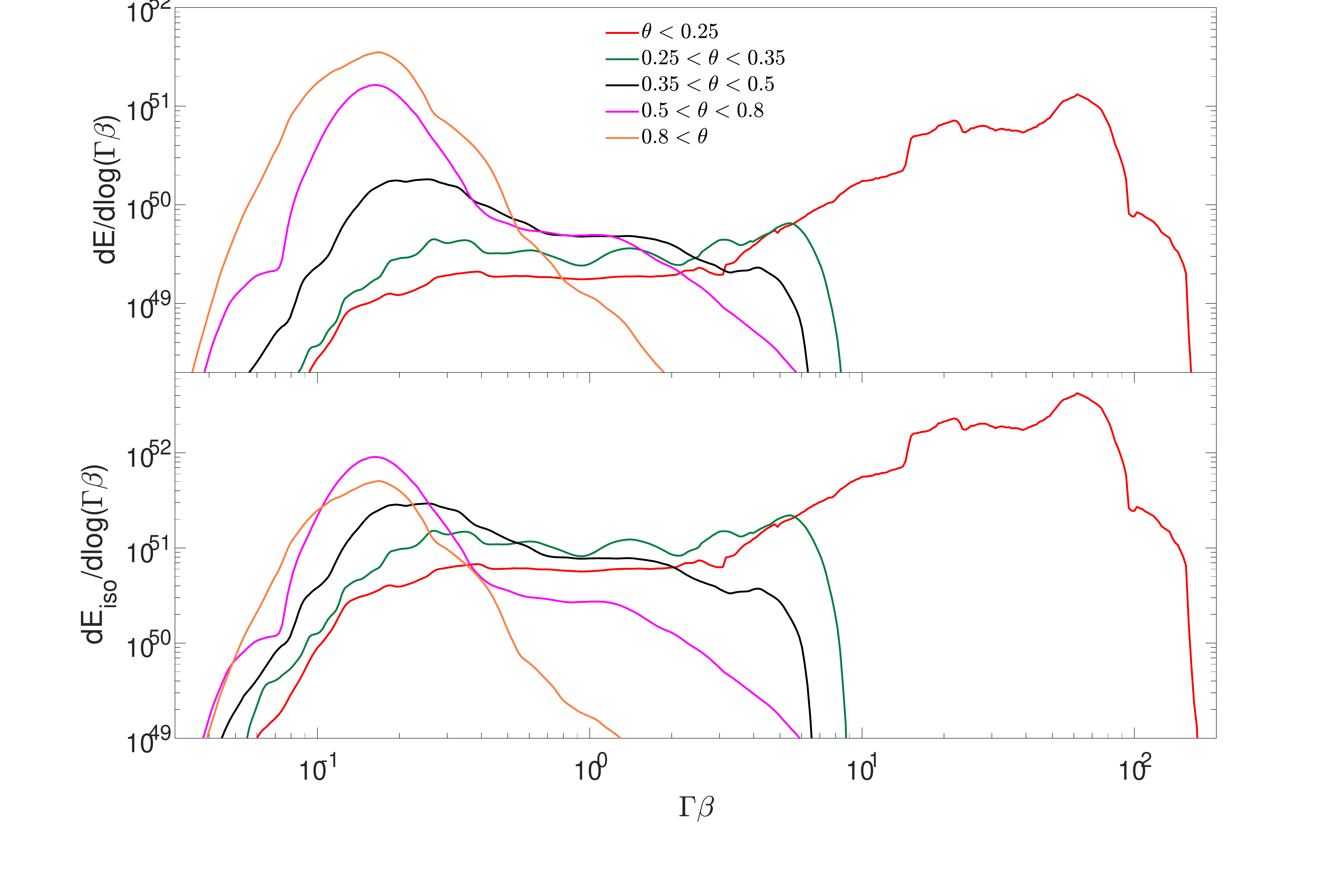}
\caption{A successful jet structure: the total energy (top panel) and isotropic equivalent energy (bottom panel) carried by the outflow per logarithmic scale of the four velocity at different angles with respect to the jet axis. The outflow is calculated by a hydrodynamic simulation of a collimated successful jet that was presented in  \citet{Gottlieb17b}. The figure shows that the outflow has a considerable radial and angular structure. In this simulation, which fits the radio data, the jet is much more energetic then the cocoon and its isotropic equivalent energy $>10^{52}$ erg/s. The cocoon dominates the energy carried by a mildly relativistic outflow  at angles between 0.25 and 0.5 rad. This component dominates most of the afterglow emission during the gradual rise.}
\label{fig:successful}%
\end{figure}

In general, the structure of the jet-cocoon outflow depends mostly on two factors. First, whether the jet is collimated or not and second, in case that it is  collimated, whether the jet is choked or not. If the jet is wide enough so it cannot be collimated then it must be choked. The jet energy is then deposited entirely into the ejecta over a large opening angle and the resulting outflow is composed only of a cocoon with a strong radial structure. The sideways expansion before and after the cocoon breakout induces also an angular structure where more energy is carried by a faster material closer to the jet axis. An example of the outflow structure (angular and radial) from a choked uncollimated jet from \cite{Gottlieb17b} is shown in figure \ref{fig:choked}. 

If the jet is collimated and the engine operates for sufficiently long  time then it will break out along with the cocoon, composing together a structured jet. An example of the outflow structure in that case taken from \cite{Gottlieb17b} is shown in figure \ref{fig:successful}. The relativistic core is composed out of jet material that was launched after the breakout and its Lorentz factor and opening angle are comparable to those that were set at the launching site. The angular structure of the cocoon is stronger than in the uncollimated case, yet there is a considerable radial structure. Most of the angular structure is dictated by the expansion after the breakout and therefore the mildly relativistic material is spreading at an angle of $\sim 0.5$ rad (see \citealt{Nakar2017}). Finally, the structure of a collimated choked jet depends on how deep it is choked. If it is not too deep then the structure is similar to that of the successful jet, only without the relativistic core. In general the energy ratio between the jet and the cocoon can assume any value, from zero if the jet is choked to $\gg 1$ if the engine works for a time that is much longer than the time it takes the jet to break out. For typical parameters we expect though, that if the jet does break out then the engine work time after the jet breaks out is comparable to its work time before the breakout and therefore the jet energy is comparable to that of the cocoon \citep{bromberg2011}.

\subsection{Is there a successful jet in GW170817 }\label{sec:jetNojet}

The question whether a jet successfully emerged from the ejecta in GW170817 is strongly linked to the connection between neutron star mergers and sGRBs \citep{Eichler1989}.  A successful jet  most likely produced a sGRB pointing away from us. 
However, the  observed gamma-rays didn't resemble an un-boosted regular sGRB and  even if GW 170817 was accompanied by a regular sGRB pointing elsewhere,  
 the prompt gamma-rays that we observed were almost certainly not  this sGRB seen off-axis,  \citep{Hallinan17,mooley2018}.
Instead they are  consistent with  a  breakout of a mildly relativistic shock driven by a cocoon into the sub-relativistic ejecta \citep{Kasliwal17,Gottlieb17b,bromberg2018,nakar2018b}.  These authors find that such a gamma-ray signal can be generated by either a choked or a successful jet.  Therefore the observed prompt gamma-rays do not provide a clue on  the fate of the jet.

We turn now to explore what can we learn from the observations on this question. The  observations up to now 
{ including the observed flattening} can be well fitted  by either a choked jet \citep{mooley2018,Gottlieb17a,nakar2018b} or a successful one \citep{mooley2018,lazzati2017c,nakar2018b,margutti2018,Troja2018,lyman2018,DAvanzo18}. As a clear jet signature was not observed so far, if the jet did break out then there are two possibilities. If the  successful jet is weak enough its contribution is sub-dominant at all times (see figure \ref{fig:schematic}). This possibility cannot be tested observationally.  It requires the jet to have much less energy than the cocoon. This is unlikely (e.g., \citealt{bromberg2011}) but possible. Alternatively, the jet could  have a comparable or a larger energy  than the cocoon. In that case the jet core dominates the emission around the peak of the light curve.

{ If there was a powerfull successful jet in GW170817 and its emission dominated the around 150 day, then at this time $\Gamma(t) \approx (\theta_{obs}-\theta_j)^{-1}$. This provides an estimate of the jet isotropic equivalent energy \citep[e.g.,][]{Granot:2002,nakar2002}:}
 \begin{equation} \label{eq:Ejiso}
E_{iso,j} \sim 3 \times 10^{52}~ {\rm erg} ~  \frac{n}{10^{-4} {\rm cm}^{-3}} \left(\frac{{\theta_{obs}-\theta}_j}{15 {\rm deg}}\right)^{-8} \left(\frac{t_{peak}}{150 {\rm day}}\right)^3 \ .
\end{equation} 
Given the constraints on the observing angle, $\theta_{obs}$ derived from the gravitational wave signal \citep{abbott2017H0} 
this requires  an extremely large isotropic equivalent jet energy, estimated by \cite{mooley2018} to be $E_{iso,j}\gtrsim 10^{52}$ erg. This is an order of magnitude estimate, and there might be some region of the phase space where a somewhat lower jet energy can be consistent with the data. Yet, all the models suggested so far in which the jet contributes to the afterglow emission at some point,  being based on jet-cocoon simulations \citep{lazzati2017c,nakar2018b,margutti2018} or on a jet structure inserted by hand \citep{lyman2018,DAvanzo18,Troja2018}, 
have a core isotropic equivalent energy $E_{iso,j}\gtrsim 10^{52}$ erg. 
For example, \cite{Troja2018} did a parameter phase phase study for a Gaussian jet structure finding that  $E_{iso,j} \gtrsim 10^{52}$ erg is necessary to fit the data and the best fit is obtained for 
 $E_{iso,j} > 10^{54}$ erg. 
However, while these models fit the data, sGRBs that are so powerful are rare. Different estimates of the sGRB luminosity function  \citep{guetta2005,nakar2006,nakar2007,guetta2009,dietz2011,petrillo2013,dAvanzo2014,wanderman2015} all find that such  bright sGRBs are less than $\sim 1\%$ of the sGRB population\footnote{Eq. \ref{eq:Ejiso} constrain $E_{iso,j}$ , while the luminosity function is given as a function of $L_{\gamma}$,  the peak $\gamma$-rays luminosity emitted within 64 ms. The link between the two depends, first, on the burst variability and duration where typically $L_{\gamma}$ is larger by a factor of a few than $E_{iso,j}/1{\rm s}$. Second, it depends on the efficiency of converting the jet energy to $\gamma$-rays. Assuming 20\% efficiency both factors cancel each other and $E_{iso} \sim 10^{52}$ erg burst corresponds to  a sGRB  with $L_\gamma \sim 10^{52}$ erg/s. } \citep[see however][who find a much brighter luminosity function]{ghirlanda2016}.

This argument suggests that a successful jet in GW170817 is unlikely. Yet, this line of reasoning does not rule it out.  First, this argument might be wrong if, for example, there is an anti-correlation between $E_{iso,j}$ and $\theta_j$ in which case bright bursts are common but rarely point towards Earth. Second, we might have been lucky and have detected a rare event. Therefore it is important to look for the jet signature in future observations.

{ As discussed below, since we did not see the off-axis emission from the core of the jet, the light curve alone will not enable us to securely identify a jet signature.} Namely, as we show in the next section (see also figure \ref{fig:schematic}), a decline in the light curve does not imply that we see the signature of a successful jet. 
Finally,  while the light curve alone may not provide the answer for the fate of the jet, other observations such as direct imaging and polarization may enable us to determine that \citep{nakar2018b}. 

\subsection{Evolution of the light curve in various scenarios} 
\label{sec:future}

\begin{figure}	
\includegraphics[width=0.48\textwidth]{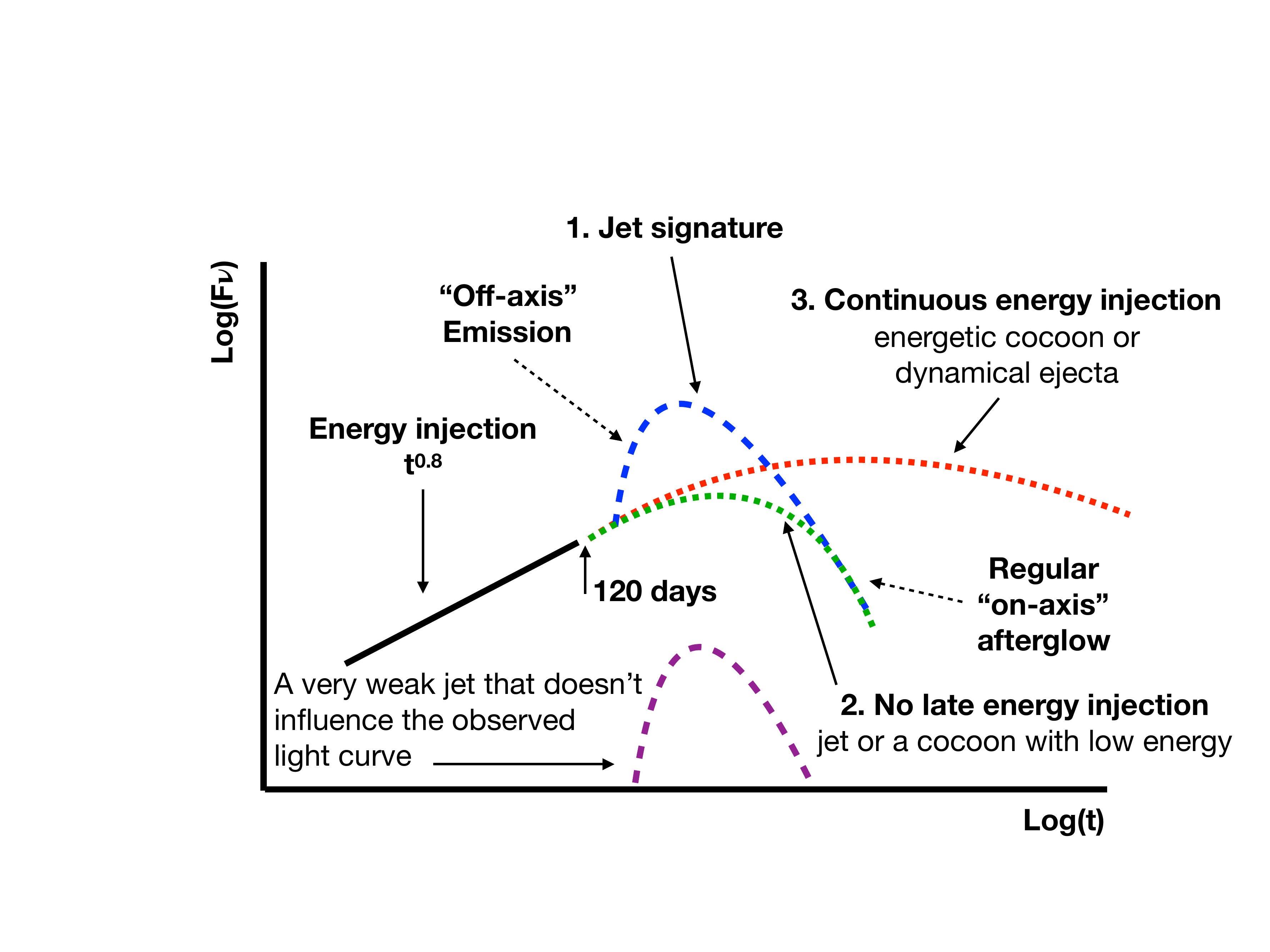}
\caption{A schematic description of the radio light curve. The solid black line depicts the $t^{0.8}$ rising signal observed during the first 100 day. (1) The dashed blue line depicts a possible ``orphan afterglow'' like jet signature. Such a signature arises if the jet is powerful enough so its off axis emission dominates before it peaks once $1/\Gamma_j \approx  (\theta_{obs}-\theta_j)$. The late decay resembles a regular on-axis afterglow. (2) The dotted green line depicts a light curve in which we observe an on-axis emission up to the time that energy injection effectively stops, which results in a gradual rise to the peak followed by a decay that resembles a regular on-axis afterglow. This scenario can be powered either by a successful jet or by a low energy cocoon formed by, most likely a collimated, choked jet.   (3) The dotted red line depicts an outflow where there is a significant energy injection also after the peak, most likely carried by at slower material. In this case the decay is shallower than a regular on-axis afterglow and it reflects the energy injection rate that continues after the peak. Such emission is not expected if there is a successful jet. It can be a result of an energetic cocoon formed by an uncollimated choked jet or a result of a fast component of the dynamical ejecta.
The purple curves shows the contribution of a very weak successful jet. Such a jet already contributed to the light curve, but its off-axis afterglow  was always below the observed light curve. }
\label{fig:schematic}%
\end{figure}

{ We turn now to consider the behavior of the light curve in  various scenarios.  
Eq. \ref{eq:Etotp} shows that the rise in the light curve requires a considerable amounts of energy injection. 
This cannot continue forever and hence the light curve levels off and then decrease
in all the models. However, the behavior before and after the peak may vary between models (figure \ref{fig:schematic}).} 		

First, we consider a successful jet that contributes to the emission. The jet's core contribution to the observed afterglow behaves like an orphan afterglow from a ``top hat'' jet. It rises rapidly and peaks once its Lorentz factor, $\Gamma_j$ satisfies $\Gamma_j (\theta_{obs}-\theta_j) \approx 1$. 
If the jet core is energetic enough its emission may dominate over the rest of the outflow already in the rapid rising phase, when its emission is still off-axis (case 1 in figure \ref{fig:schematic}). Such a rapid rise provides a direct evidence for the existence of a successful jet. { Unfortunately, this signature was not seen on the light curve so far.} Alternatively, the jet may dominate only once its emission becomes on-axis. In that case the light curve will continue its steady rise to a peak, which is seen when the core dominates (case 2 in figure \ref{fig:schematic}). In both cases there is no more energy injection after the core of the jet is fully on-axis and therefore the light curve following the peak is expected to decay at a rate that is similar to a regular on-axis GRB afterglow after the jet break, i.e, roughly $F_\nu \propto t^{-p}$ in the relativistic regime and $t^{-(15p-10)/21}$ in the newtonian regime (since the outflow is mildly relativistic the declines will be between these two values). The peak in that case enable us to measure the total energy in the jet and together with the upper limit on the observing angle from the GW signal, also to constrain its initial opening angle and isotropic equivalent energy. Thus, if the afterglow is dominated by a successful jet then the light curve is predicted to decay rapidly in the near future. Note, however, that while case 1 (which was not observed), is unique to a successful jet, case 2 is not (see below) and therefore the light curve alone, even if it decays rapidly following the peak, does not provide a direct evidence for a successful jet. 

Second, we consider the emission from a cocoon with a choked jet. In that case the rise to the peak is relatively smooth with possible moderate changes in the slope. After the peak, the evolution depends on the exact cocoon structure. If the peak is observed after most of the cocoon energy is already deposited in the blast wave, then the decay will resemble a regular on-axis afterglow (case 2). This case is more likely if the jet is collimated before it is choked.  If, instead, energy will still be  injected into the blast wave after the peak (most likely by a radial structure in that case), then the decay will be shallower than that of an on-axis afterglow (case 3 in figure \ref{fig:schematic}). This case is more likely if the jet is not collimated. Thus, for a choke jet we expect to see either case 2 or case 3,

Third,  if the afterglow is powered by the dynamical ejecta, with no significant contribution from a jet or a cocoon, then energy injection will continue until we see the sub-relativistic macronova ejecta in which case the light curve will follow case 3 in figure \ref{fig:schematic}.

{ The recent flattening and possible turnover of the light curve suggest that  in GW170817 we have  either case 2 or 3. The ultimate difference between the two cases can be determined only in future observations that will show the rate in which the light curve declines. If the light curve will show a rapid decay in the near future (e.g., case 2), then it will provide tight constraints on the properties of the jet (up to uncertainty due to unknown ISM density and microphysics parameters). First, the jet's total energy is a few times $10^{49}$ erg, regardless of whether it was choked or successful, and it was collimated during its propagation through the ejecta. Second, if the jet was successful, then its isotropic equivalent energy is $\sim 10^{52}$ erg, which implies an initial opening angle of about $0.1$ rad and an observing angle of $0.4-0.6$ rad. These properties are shared by all the published models that fit the data with a dominant successful jet \citep{lazzati2017c,nakar2018b,margutti2018,lyman2018,Troja2018}. A gradual decay (or even  a renewed  rise),  will rule out a successful jet as the origin of the afterglow and will support an uncollimated choked jet or a dynamical ejecta as the afterglow source (these two models can be differentiated using their different prediction for the cooling frequency; \citealt{hotokezaka2018b}).
Nevertheless, even in that case we won't be able to rule out a very weak jet whose contribution to the afterglow was insignificant at all times.} 

Finally, we  note that the very long term evolution is known. 
The macronova observations revealed a significant mass, $\sim 0.01 M_\odot$, within the dynamical ejecta moving at $\sim 0.3$c. 
This will eventually lead to a classical Newtonian radio flare \citep{nakar2011} that will peak on a time scale of years (or even dozens of years), depending on the external density,  most likely at a lower or comparable luminosity than the current signal. 

\section{Conclusions} 
\label{sec:conc}

We have explored the  implications of the unique radio to X-ray emission that followed the gravitational waves from GW170817.  This emission had a single power-law spectrum and a gradually monotonously increasing light curve over a decade  in time { until it started flattening around 115 days}. Such a spectrum and light curve were never observed in any GRB afterglow.  Still this emission can be interpreted within the context of a blast wave propagating into a constant density medium.
The observed spectrum indicates that we are within a single spectral regime  that is below the cooling frequency and above the typical synchrotron and self absorption frequencies from the first observations at day 9 until at least day 150.

Defining  off-axis  as emission from matter moving towards the observer  at  an angle $> 1/\Gamma$ with respect to the line of sight and  on-axis as emission  from matter moving {\it at the time of the emission} within an angle $<1/\Gamma$ (the latter matter could have moved at a larger Lorentz factor and/or a larger angle at an earlier time),  we show that an off-axis emission must rise faster than $t^3$. This is clearly inconsistent with the observed slow increase in the afterglow flux.   We conclude, therefore, that so far we have observed at all times  on-axis emission from matter moving towards us. 

As on-axis emission within this spectral range is extremely sensitive to the Lorentz factor we  tightly constrain it: $1.5 < \Gamma < 7$  on day 10 and a slow decline at later time: $\Gamma \propto t^{-0.21}$.  Thus, the emitting matter is moving towards us at a mildly relativistic velocity. The monotonous increase in the light curve must arise from energy injection. The isotropic equivalent energy in the observed region is $E_{iso} \sim  10^{50}~ (t/150~ {\rm days})^{1.3}$, while the total energy in the observed region  is $E_{obs} \sim 10^{49}~ (t/100~ {\rm days})^{1.73}$ erg. This energy injection can arise from  slower material moving behind the blast wave or from matter moving at a different angle that has slowed down so that we have entered its $1/\Gamma$ beam. This emission is most likely the mildly relativistic component of the radio flare predicted long ago to accompany neutron star mergers \citep{nakar2011}.

The energy injection implies that the emitting region has a structure,  either radial or angular or both. A moderately rising afterglow was predicted before these observations in some models where the outflow has a structure. This includes  jet-cocoon models \citep{Gottlieb17b} as well as orphan afterglows where the jet structure was arbitrary assumed \citep{granot2005,lamb2017}. While there is a lot of freedom in the distribution of this structure, the only physical mechanism that we are aware of that naturally generates it, is the propagation of a relativistic jet within the ejecta surrounding the merger. The jet-ejecta interaction produces a hot cocoon that, for typical parameters, has the right energy and Lorentz factor \citep{Gottlieb17b}.
In fact the existence of such a cocoon within neutron star mergers was predicted even before this event \citep{Nakar2017,lazzati2017a,Gottlieb17a}.
Such a cocoon forms whenever a jet propagates within a surrounding medium, regardless of whether the jet is   choked  within the ejecta \citep{mooley2018,Gottlieb17a,nakar2018b} or crosses it successfully \citep{mooley2018,lazzati2017c,Gottlieb17a,nakar2018b,margutti2018}.  This last structure of a successful jet and its cocoon is sometimes called a ``structured jet'', although the origin of the structure is not necessarily the same as jets whose ad-hoc structure is inserted by hand (e.g., \citealt{lyman2018,DAvanzo18}).  An advantage of the cocoon model is that it also naturally explains the origin of the prompt gamma-rays \citep{Kasliwal17,Gottlieb17b,bromberg2018,nakar2018b}, that do not seem to be a regular sGRB prompt emission observed off-axis \citep{Hallinan17,mooley2018,Gottlieb17b}. Finally, If the dynamical ejecta have a fast tail, which is roughly a spherical outflow with a radial structure, then it can also produce the observed radio to X-ray emission \citep{mooley2018,hotokezaka2018b}, but it is not clear what is the origin of the prompt gamma-rays in this case.  

{ The end of the rise in the light curve around day 150 indicates on a decrease in the rate that energy is injected into the observed region.  If the energy injection has stopped entirely (case 2 in figure \ref{fig:schematic}) then the peak will be followed by a rapid decay (faster than at least $t^{-1.2}$), while a shallower decline will indicate that energy injection continues  (case 3 in figure \ref{fig:schematic}). The isotropic equivalent energy on day 150 is $\sim 10^{50}$ erg and the total energy in the observed region is $\sim 2 \times 10^{49}$ erg (these values can vary by about an order of magnitude depending on the uncertain ISM density and microphysical parameters). If the energy injection has stopped it implies that this is the total energy of the jet, and that the jet was collimated regardless of whether it was choked or successful. If future evolution will indicate that energy injection continues, then it will support a choked uncollimated jet or a the quasi-spherical merger ejecta as the source of the afterglow. In that latter case an interesting possible diagnostic will be the detection of the cooling break moving into the X-ray band. This can happen if the external density is on the high side of the parameter phase space and the outflow velocity is on the low side. This is expected if the afterglow is powered by the dynamical ejecta \citep{hotokezaka2018b}. If observed, it will provide an estimate to one of the least known parameters and enable us to firmly constrain the energy involved. Another interesting possibility is that radio measurements will enable us to resolve the ejecta and/or  measure the polarization. Both will provide information that may enable us to distinguish between a choked and a successful jet \citep{nakar2018b}.}

While extremely illuminating, the radio to X-ray emission did not answer so far one of the most interesting questions concerning GW170817: was it accompanied by a regular sGRB pointing elsewhere. This  most likely has happened if the jet successfully penetrated the ejecta.  In principle such an sGRB jet could have had a strong bright orphan afterglow that would have over-shine the emission of the rest of the ejecta. This would have been clearly identified. Unfortunately, so far this is not the case. This implies that either the jet is too weak and its afterglow was always fainter than the emission of other components, or that the jet core is very powerful (with $E_{iso,j} \gtrsim 10^{52}$ erg). Clearly it is also possible, and we argue that it is even likely, that in this particular merger the jet was choked in the ejecta having no sGRB at all.

Future observations of many mergers, expected once LIGO/Virgo resume operation, or even before that if a macronova will be identified optically, will enable us to explore these questions statistically.  A sample of several dozens of mergers viewed from different angles, including a few with an sGRB pointing to us,  and their afterglow could enable us to estimate what is the angular structure of the outflow and from this the properties of the jets. 
 While independent evidence suggest that a large fraction of jets in merger events are choked \citep{Moharana2017}, such data will provide a good estimates of the statistics of choked vs. successful jets. Confronting this with the statistics of merger rates vs sGRB rates we can obtain independent estimates of sGRB beaming.

We thank Ore Gottlieb, Kenta Hotokezaka, Gregg Hallinan and Mansi Kasliwal  for helpful discussions and Om Sharan Salafia for helpful comments. This research was supported by the I-Core center of excellence of the CHE-ISF.  EN was partially supported by an ISF grant (1114/17). TP was partially supported by an advanced ERC grant TReX and by a grant from the Templeton foundation.

\end{document}